\documentclass[final,3p,times,twocolumn]{elsarticle}

\usepackage{amssymb}
\usepackage{amsmath}
\usepackage[T1]{fontenc} 

\journal{Radiation Measurements}

\begin{document}

\begin{frontmatter}



\title{Performance of a plastic scintillator developed using styrene monomer polymerization}


\author{A. Sadigov
\fnref{fn1,fn2}}

\author{F. Ahmadov
\fnref{fn1,fn2}}
\ead{farid081211@gmail.com}

\author{G. Ahmadov
\fnref{fn1,fn2,fn3,fn4}}

\author{E. Aksu
\fnref{fn5}}

\author{D. Berikov
\fnref{fn4,fn6}}

\author{S. Nuruyev
\fnref{fn1,fn4}}

\author{R. Akbarov
\fnref{fn1,fn2,fn4}}

\author{M. Holik
\fnref{fn7,fn8}}

\author{J. Nagiyev
\fnref{fn2}}

\author{S. G\"urb\"uz G\"uner
\fnref{fn5}}

\author{A. Mammadli
\fnref{fn1}}

\author{N. Suleymanova
\fnref{fn1}}

\author{C. Abbasova
\fnref{fn1}}

\author{S. Melikova
\fnref{fn1}}

\author{E. Yilmaz
\fnref{fn9}}

\author{O. Tagiyev
\fnref{fn10}}

\author{S. Lyubchyk
\fnref{fn11}}

\author{Z. Sadygov
\fnref{fn1,fn2,fn3}}

\affiliation[fn1]{organization={Institute of Radiation Problems under Ministry of Science and Education},
            addressline={B.Vahabzade str. 9}, 
            city={Baku},
            postcode={AZ1143}, 
            country={Azerbaijan}}

\affiliation[fn2]{organization={Innovation and Digital Development Agency Nuclear Research Department},
            addressline={Gobu str. 20th km of Baku-Shamakhi Highway}, 
            city={Baku},
            postcode={AZ1021}, 
            country={Azerbaijan}}
            
\affiliation[fn3]{organization={Innovative Electronics and Detectors LLC},
            addressline={Badamdard STQ-1}, 
            city={Baku},
            postcode={AZ1021}, 
            country={Azerbaijan}}
            
\affiliation[fn4]{organization={Joint Institute for Nuclear Research},
            addressline={Joliot-Curie 6}, 
            city={Dubna},
            postcode={141980}, 
            country={Russia}}
            
\affiliation[fn5]{organization={T\"urkiye Energy, Nuclear and Mineral Research Agency, Nuclear Energy Research Institution},
            addressline={Saray, Atom Cd. 27}, 
            city={Ankara},
            postcode={06980}, 
            country={T\"urkiye}}

\affiliation[fn6]{organization={Institute of Nuclear Physics of the Ministry of Energy of the Republic of Kazakhstan},
            addressline={Ibragimova 1}, 
            city={Almaty},
            postcode={050032}, 
            country={Kazakhstan}}
        
\affiliation[fn7]{organization={Faculty of Electrical Engineering, UWB in Pilsen},
            addressline={Univerzitni 2795/26}, 
            city={Pilsen},
            postcode={306 14 }, 
            country={Czech Republic}}
            
\affiliation[fn8]{organization={Institute of Experimental and Applied Physics, CTU in Prague},
            addressline={Husova 240/5}, 
            city={Prague},
            postcode={110 00 }, 
            country={Czech Republic}}            
            
\affiliation[fn9]{organization={Department of Physics, Bolu Abant Izzet Baysal University},
            addressline={G\"olk\"oy Kamp\"us\"u}, 
            city={Bolu},
            postcode={14280}, 
            country={T\"urkiye}}                

\affiliation[fn10]{organization={Institute of Physics -- Ministry of Science and Education},
            addressline={G.Javid ave. 131}, 
            city={Baku},
            postcode={AZ1143}, 
            country={Azerbaijan}}                            
            
\affiliation[fn11]{organization={Universidade Lusofona},
            addressline={Campo Grande 376}, 
            city={Lisbon},
            postcode={1749-024}, 
            country={Portugal}}

\begin{abstract}

This paper presents a newly developed plastic scintillator produced in collaboration with T\"urkiye Energy, Nuclear and Mineral Research Agency (TENMAK). The scintillator is manufactured using thermal polymerization of commercially available styrene monomer. The absorption spectrum of the scintillator exhibited two absorption bands at 225 nm and 340 nm, with an absorption edge observed at 410 nm. The wavelength of the emitted light was measured in the range of 400-800 nm, with a maximum intensity at 427 nm. Monoenergetic electrons from the $^{137}$Cs source were used to evaluate the characteristics of the new scintillator, particularly its light yield. As the light readout the MAPD-3NM type silicon photomultiplier array (4 $\times$ 4) with an active area of 15 $\times$ 15 mm$^2$, assembled using single MAPDs with an active area of 3.7 $\times$ 3.7 mm$^2$, was used. The light yield of the scintillator was determined to be 6134 photons/MeV. In addition, the efficiency of the scintillator for gamma rays with an energy of 662 keV was found out to be approximately 1.8 $\%$. A CmBe neutron source was employed to evaluate  its fast neutron detection performance. However, neutron/gamma discrimination using pulse shape discrimination (charge integration) method was not observed. The results demonstrate the potential of a newly produced plastic scintillator for various applications, particularly in radiation monitoring and detection systems.

\end{abstract}




\end{frontmatter}


\section{Introduction}
\label{sec:level1}

Plastic scintillators with photo sensors have found wide application in high-energy physics, space exploration, medical diagnostics, and security systems \cite{bib1,bib2,bib3,bib4,bib5}. The widespread use of plastic scintillators can be explained by advances in their manufacturing technology as well as a number of valuable properties of the scintillator itself, such as short decay time, high radiation resistance, operating temperature, resistance to mechanical stress, etc. Moreover, the ability to produce plastic scintillators in various shapes and sizes makes them suitable  converters of ionizing energy into visible light, rendering them suitable for numerous experiments \cite{bib4,bib5,bib6}. In light of these factors, the production of plastic scintillators continues to be of utmost significance, driving ongoing research and development efforts aimed at further improving their performance, cost-effectiveness, and compatibility with emerging detection systems.
      
One notable advantage of plastic scintillators is their exceptionally short decay time, which makes them indispensable for experiments that require precise time-of-flight measurements. Additionally, due to the low atomic number, they are effective in detecting charged particles and neutrons \cite{bib6}. The interaction of gamma rays with plastic scintillators is primarily related to their atomic number (Z) and energy of gamma ray ($E_{\gamma}$), resulting in a minimal probability of the photoelectric effect ($\sigma_{pe.} \sim \frac{Z^{n}}{E_{\gamma}^{3.5}}$). Consequently, the energy of gamma rays in plastic scintillators is typically determined by the Compton edge ($\sigma_{comp.} \sim const. \cdot Z$) and the maximum energy of a Compton electron can be calculated using the following formula: 

\begin{equation}
\label{eqn:eq01}
E_{e} =\frac{2 \cdot E_{\gamma}^{2}}{2 \cdot E_{\gamma} + m_{e} \cdot c^{2}}
\end{equation}
where $E_{e}$ - the maximum energy of a Compton electron, $m_{e}$ - the mass of an electron and $c$ - the speed of light \cite{bib7,bib8}. Table \ref{tab:1} lists the maximum energy of Compton electrons produced by various gamma rays.

\begin{table*}[tp]
\caption{\label{tab:1}The maximum energy of Compton electrons produced by various gamma rays.}
\vspace*{3mm}
\centering
\begin{tabular}{c c c}
\hline
\hline \\
Radionuclide & Energy of gamma-rays (keV) & Maximum energy of Compton electrons (keV)\\
\hline
$^{241}$Am & 59.6 & 11.27 \\[0.05cm]
\hline
$^{133}$Ba & 81, 276, 303, 356, 384  & 19.5, 143.3, 164.4, 207.2, 230.6 \\[0.05cm]
\hline
$^{137}$Cs & 661.6 & 477.65 \\[0.05cm]
\hline
$^{60}$Co & 1173, 1332 &  963.2, 1117.6\\[0.05cm]
\hline
\hline
\end{tabular}
\end{table*}

When ionizing radiation interacts  with the scintillator material, the scintillator molecules get excited, and then they return to their ground state with emitting  scintillation photons. All absorbed energy of ionizing radiation in the scintillator is not converted to visible scintillation light. Ideally, the number of photons generated should  increase linearly with the incident radiation energy, although deviations from linearity can occur depending on the specific characteristics of the scintillation material. The light yield of the scintillator is calculated as follows \cite{bib4}:

\begin{equation}
\label{eqn:eq02}
Y =\frac{E_{dep.} \cdot \eta}{E(\lambda)}
\end{equation}

where $E_{dep.}$ - deposited energy of ionizing radiation, $\eta$ - efficiency of scintillator and  $E(\lambda)$ - the energy of an emitted photon. The light yield of the scintillator depends on the scintillation material, the type of incident particles, the energy of particles and the temperature \cite{bib5}.

In scintillation detectors, the phenomenon of quenching plays a significant role in the detection and measurement of ionizing radiation. Quenching occurs due to various mechanisms associated with the interaction of radiation with the scintillator material. Scintillator quenching is generally described by different models (Birks, Yoshida, Voltz) \cite{bib9, bib9_2}. In these models are taken into account affect electronic stopping, nuclear stopping, the singlet quenching, the triplet quenching and others effects. The luminescence yield per unit length in an organic scintillator ($\frac{dY}{dx}$) can be calculated given the ionization density  ($\frac{dE}{dx}$), the energy transfer probability (k), a normalization factor (s), the constant of proportionality (B) to determine the number of damaged molecules, the energy emitted as light (Y), and particle energy dissipated in the scintillator (E)

\begin{equation}
\label{eqn:eq03}
\frac{dY}{dx} =\frac {s \cdot \frac{dE}{dx}}{1+kB \cdot \frac{dE}{dx}}.
\end{equation}

Different types of particles, such as electrons, alpha particles,  heavy ions, can exhibit distinct quenching behaviors. Quenching of the light yield is related to the energy transfer to damaged molecules, which do not convert the received energy into scintillation photons. These features, which leads to variations in the shape of the detected pulses, can be employed to distinguish  different types of ionizing radiation \cite {bib5,bib9}. It is possible to discriminate particles type by analyzing pulse shapes, thereby enhancing the capabilities of radiation detection systems. This performance of plastic scintillation detectors  allows use them in PSD experiments \cite{bib4}. 

The light yield of the using plastic scintillators  mainly changes in the range of 3000 - 20000 photons/MeV e.g : EJ-232-0.5$\%$ (2900 photons/MeV), EJ-256-5$\%$ (5200 photons/MeV), EJ-240G (6300 photons/MeV), EJ-254-5$\%$ (7500 photons/MeV), EJ-276G (8000 photons/MeV), EJ-290 (9000 photons/MeV), EJ-212 (10000 photons/MeV) and others \cite {bib10, bib11}.

The main objective of this study is to investigate the parameters and performance of the plastic scintillator produced by TENMAK. The absorption and photoluminescence performance of this scintillator were measured, and the light yield was determined using the light output of the commercial standard $LaBr_{3}(Ce)$ scintillator by Epic Crystal. To gain insight into the scintillation mechanism, the optical performance, scintillation responses, and pulse shape discrimination (PSD) performance on $\gamma$-rays, neutrons, and beta particles were investigated using MAPD-3NM type silicon photomultiplier as a light readout and various sources including $^{133}$Ba, $^{137}$Cs, and $^{60}$Co, $^{90}$Sr, and $^{244}$CmBe neutron source.

\section{\label{sec:level2}Experimental}

In this study, we used a polystyrene plastic scintillator with  a rectangular shape and dimensions of 15 $\times$ 15 $\times$ 50 mm$^3$ to convert the energy of ionizing radiation (gamma ray and beta particle) to visible light pulse. The plastic scintillator was produced through  thermal polymerization of commercially available styrene monomer. The production process involved several steps. Firstly, purified styrene monomer was heated to 120 $^{\circ}$C over a period of 5 - 10 hours. It was maintained at this  temperature for 3 days to ensure complete polymerization. Afterward, the material was slowly cooled to a temperature below the glass transition temperature of polystyrene (T$_g$ = 100 $^{\circ}$C) and underwent  annealing to relieve internal stresses and avoid cracking. Finally, the scintillator  was cut into the desired shape from the fabricated piece, and the surfaces were polished to achieve a mirror-like finish. To assess the properties of the scintillator, including scintillation properties and density, measurements were conducted.

\begin{figure}[tp]
\centering
\includegraphics[width=.5\textwidth]{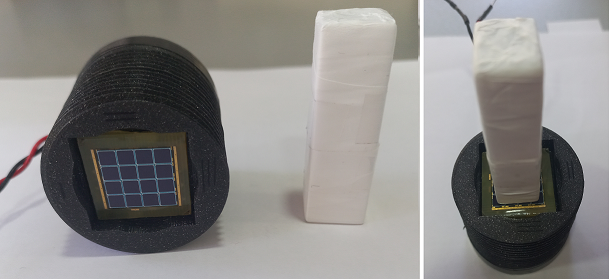}
\caption{Photo of the electronic system and the detector based on MAPD and plastic scintillator.}
\label{figure:1}
\end{figure}

Silicon photomultiplier (SiPM) array (type MAPD-3NM) was used as a detector to read out scintillation photons produced by ionizing radiation in the scintillator (figure \ref{figure:1} left). MAPD-type SiPMs have been designed and developed by our research group, employing a deeply buried pixel design. This design incorporates a double n-p-n-p junction with micro-well structures located below the surface. The structure of these SiPMs does not include conventional quenching resistors, in contrast to SiPMs produced using standard surface-pixel technology. Triggered avalanche is quenched through the utilization of specially designed potential barriers. For a comprehensive understanding of the structure and operational principles of MAPD-type SiPMs, refer to \cite {bib12, bib13, bib14, bib15, bib16, bib17}. Additional insights into the performance of MAPD as both an individual device and an array, with various scintillation materials, can be found in references \cite {bib18, bib19, bib20, bib21, bib22}. 

The plastic scintillator was covered with teflon layers and connected to the photodiode via optical glue (figure \ref{figure:1} right). The SiPM array used consisted of 16 elements, with each element being connected in parallel. A voltage was applied to SiPM array via red wire, which was linked to the cathode and the signal was taken from the black wire, connected to the anode (figure \ref{figure:1} left). The SiPM array has the following parameters: size - 15 $\times$ 15 mm$^2$, operation voltage - 55.2 V, capacitance - 155 pF, breakdown voltage - 51 V, average gain - 2 $\times$ $10^{5}$, photon detection efficiency (PDE)- 30-35 $\%$. 

\begin{figure}[tp]
\centering
\includegraphics[width=.5\textwidth]{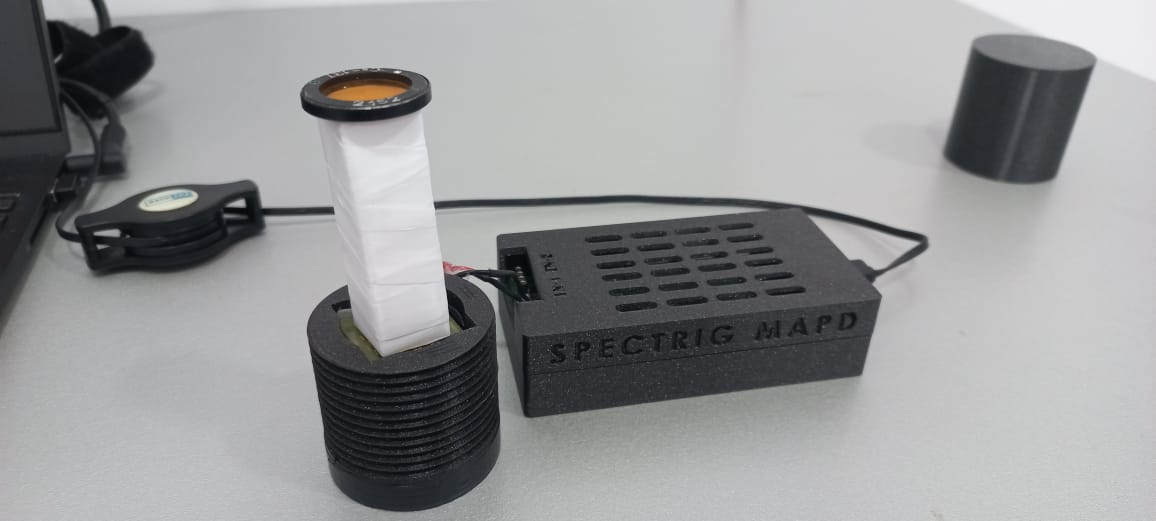}
\caption{Photo of the experimental setup.}
\label{figure:2}
\end{figure}

The ionizing radiation measurements were carried out with the Spectrig MAPD device. Detailed information about the Spectrig MAPD device can be found in \cite{bib21, bib24, bib25}. During measurement with the Spectrig MAPD, the following parameters were selected: gate width - 180 ns, variable gain - 1 - 15 dB, threshold - 43 mV, bias voltage - 55.2 V and measurement time - 200 sec and 500 sec. All measurements were carried out at a temperature of 22 $^{\circ}$C. Figure \ref{figure:2} presents a photo of the experimental setup.
 
\section{Result and discussion}
\label{sec_3}

\begin{figure}[tp]
\centering
\includegraphics[width=.5\textwidth]{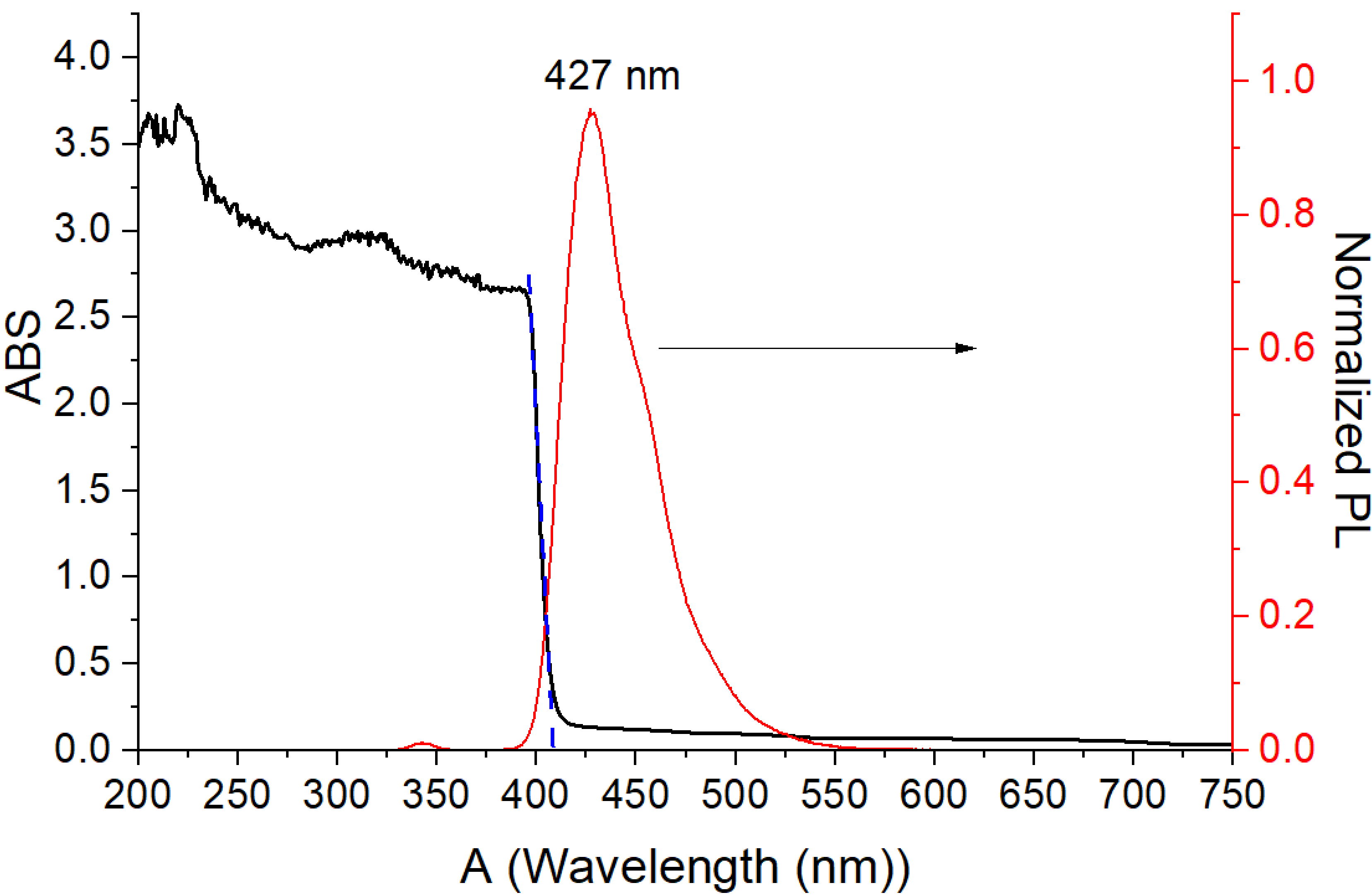}
\caption{The absorption and photoluminescence spectra of the plastic scintillator.}
\label{figure:3}
\end{figure}

Figure \ref{figure:3} shows the absorption and photoluminescence spectra of the plastic scintillators. The absorption and photoluminescence spectra were recorded by LS 55 Fluorescence Spectrometer (Perkin Elmer) \cite{bib22} and Cary 50Scan UV-Vis Spectrophotometer (Varian) \cite{bib23}, respectively. The absorption spectrum was measured in the wavelength range of 200-800 nm. 

The absorption spectrum exhibits two absorption bands at 225 and 340 nm, with an  absorption edge observed at 410 nm (figure \ref{figure:3}). As shown in figure, the transmission of light through the scintillators decreases in the region of 400-800 nm, with the transmission loss increasing as the radiation wavelength increases up to 550 nm. Beyond 550 nm, no significant changes are observed in the spectra. The photoluminescence spectrum of the scintillator was measured using a xenon lamp with a continuous spectrum ranging from 230 to 50 nm at room temperature. The spectrum exhibits a wide band covering the wavelength range of 350-550 nm,  with a  maximum corresponding to ~ 427 nm (figure \ref {figure:3}).

\begin{figure}[tp]
\centering
\includegraphics[width=.5\textwidth]{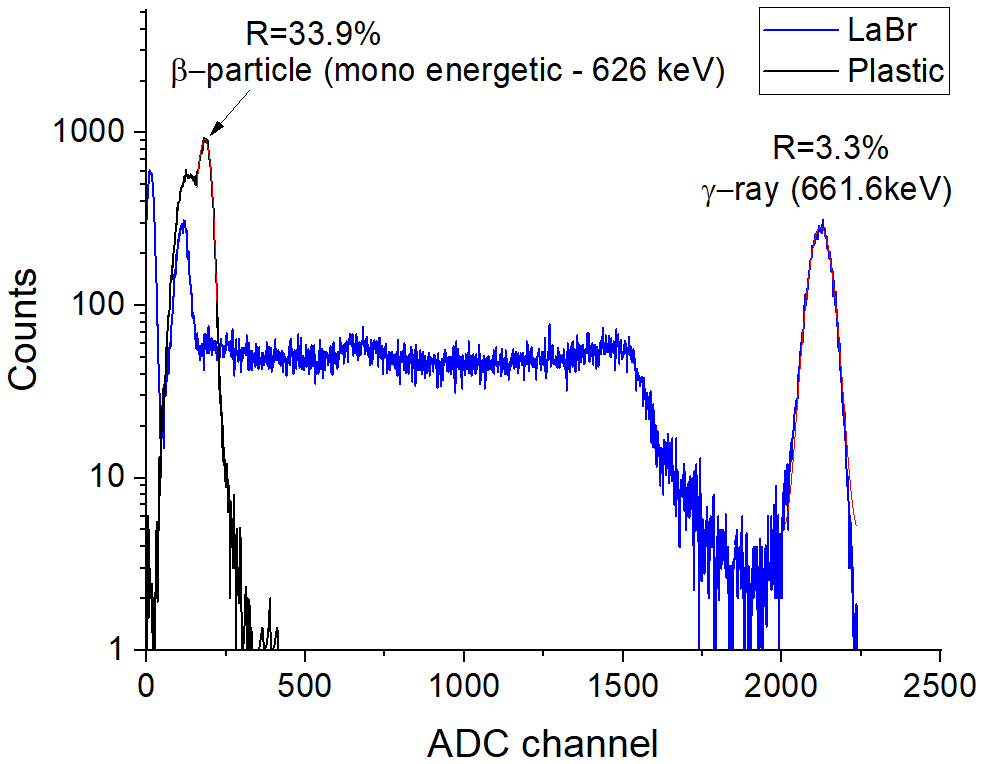}
\caption{Energy spectra of $^{137}$Cs measured with the SiPM array + $LaBr_{3}(Ce)$ and the SiPM array+plastic scintillator (the variable gain was selected 1 dB).}
\label{figure:4}
\end{figure}

Energy spectra of $^{137}$Cs measured with the SiPM array + $LaBr_{3}(Ce)$ \cite {bib24, bib25} and the SiPM array + plastic scintillator were shown in figure \ref {figure:4}. The purpose of using the $LaBr_{3}(Ce)$ scintillator was to calculate the light yield of the plastic scintillator. It is well known that $^{137}$Cs decays by beta emission to a metastable state ($^{137m}$Ba), and directly the ground state of $^{137}$Ba, with both cases resulting in the emission of gamma rays ($\sim$ 662 keV). In the first case, there is approximately a 9.6 $\%$ probability that the emitted gamma rays are captured by K- shell electrons of $^{137}$Ba, generating a monoenergetic beta particle with an energy of 626 keV. The gamma ray losses about 36 keV of energy due to the binding energy of $^{137}$Ba's K-shell electrons \cite{bib3}. 

The amplitude of the signal can be calculated as $A = PDE \times M \times N$, where N (related to scintillator) is the number of scintillation photons produced by 1 MeV of ionization radiation (or light yield of scintillator), M is the gain of SiPM, and PDE (related to SiPM) is the photon detection efficiency, which depends on the over-voltage and wavelength of the photon \cite {bib5}. Considering that the photon detection efficiency and gain of the SiPM are about the same for both emission wavelengths of scintillators, the amplitude of the signal will depend on the light yield of the scintillators. The light yield of the $LaBr_{3}(Ce)$ scintillator is determined to be 68000 photons per MeV of incident energy, corresponding to approximately 45016 photons for the incident gamma ray with an energy of 662 keV \cite {bib24, bib25}. The detected gamma ray signal with an energy of 662 keV, using the $LaBr_{3}(Ce)$ scintillator, has an amplitude of 2122 ADC channel. In this case, 626 keV energy will approximately correspond to the 2006.6 ADC channel, and the number of scintillation photons will be approximately 42568 photons. Conversely, when using the plastic scintillator, the amplitude of the detected mono-energetic beta particle signal with an energy of 626 keV is measured to be 181 ADC channel. Considering this information one can determine the light yield of the plastic scintillator using the following ratio:
\begin{equation}
\label{eqn:eq04}
\frac{A_{LaBr_{3}(Ce)} (626 \, keV)}{N_{LaBr_{3}(Ce)} (626 \, keV)} = \frac{A_{pl} (626 \, keV)}{N_{pl} (626 \, keV)}
\end{equation}

\begin{equation}
\label{eqn:eq05}
N_{pl} \, (626 keV)= N_{LaBr_{3}(Ce)} \, (626 \, keV) \times \frac{A_{pl} \, (626 \, keV)} {A_{LaBr_{3}(Ce)}}=42568 \, ph. \times \frac{181 \, ADC} {2006.6 \, ADC} = 3840 \, ph.
\end{equation}

In the case of the plastic scintillator, the number of scintillation photons corresponding to the incident gamma ray with an energy of 626 keV was found to be 3840. The light yield of the plastic scintillator for 1 MeV of incident energy was determined to be 6134 photons. It is important to note that in the given calculation, factors such as  light loss due to reflector materials, linearity of scintillators and the dependence of PDE on wavelength were not taken into account. Although in the works of the authors \cite{bib26_2, bib26_3} on the study of the non-proportionality of electron response in scintillators, it can be seen that for some plastic scintillators in the region of 626 keV, the electron response and relative light yield are approximately 100 $\%$. The efficiency ($\eta$) of the scintillator can be calculated as the following formula \cite {bib4}:

\begin{equation}
\label{eqn:eq06}
\eta =\frac{E_{ph}(\lambda) \times N_{ph}}{E_{\gamma}}
\end{equation}

where $E_{ph}$ - corresponding energy to a scintillation photon, $N_{ph}$ - the number of scintillation photons produced by 1 MeV of ionization radiation, and E - the energy of the absorbed gamma ray. In this experiment,  the energy corresponding to 427 nm was measured to be 2.9 eV, the light yield was determined to be 6134 photons/MeV. The obtained efficiency of the scintillator was approximately 1.8 $\%$. The energy resolution of the monoenergetic electron (626 keV) measured with the plastic scintillator was 33.9 $\%$. On the other hand, in the case of the measurement with $LaBr_{3}(Ce)$, the energy resolution for 662 keV gamma rays was 3.3 $\%$. In these measurements, a variable gain of MAPD Spectrig was selected 1 dB.

\begin{figure}[tp]
\centering
\includegraphics[width=.5\textwidth]{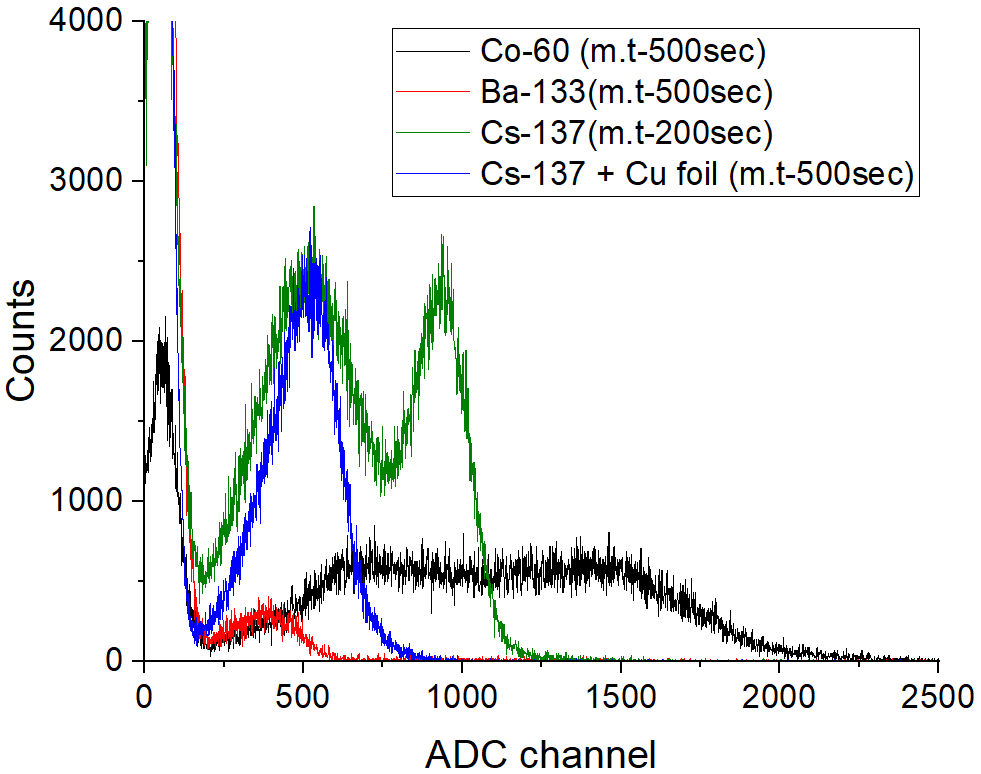}
\caption{Energy spectra of $^{133}$Ba, $^{137}$Cs and $^{60}$Co measured with the SiPM array + plastic scintillators (the variable gain was selected 15 dB).}
\label{figure:5}
\end{figure}

Figure \ref{figure:5} shows the energy spectra of $^{133}$Ba, $^{137}$Cs and $^{60}$Co measured using the Spectrig MAPD with the variable gain of 15 dB. To distinguish beta particles from gamma rays emitted from $^{137}$Cs, a copper foil with a thickness of 1 mm was placed between the scintillator and  the source. Consequently, the monoenergetic (770-1200 ADC channel) and low-energy parts (200-730 ADC channel) of the beta particles were effectively absorbed by the foil, resulting in the observation of the Compton continuum and edge in the spectrum. The measurement time was selected based on the activity of the  sources. The measurement time was selected 200 seconds for $^{137}$Cs source, while 500 seconds for $^{133}$Ba, $^{137}$Cs with the copper foil and $^{60}$Co.

\begin{figure}[tp]
\centering
\includegraphics[width=.5\textwidth]{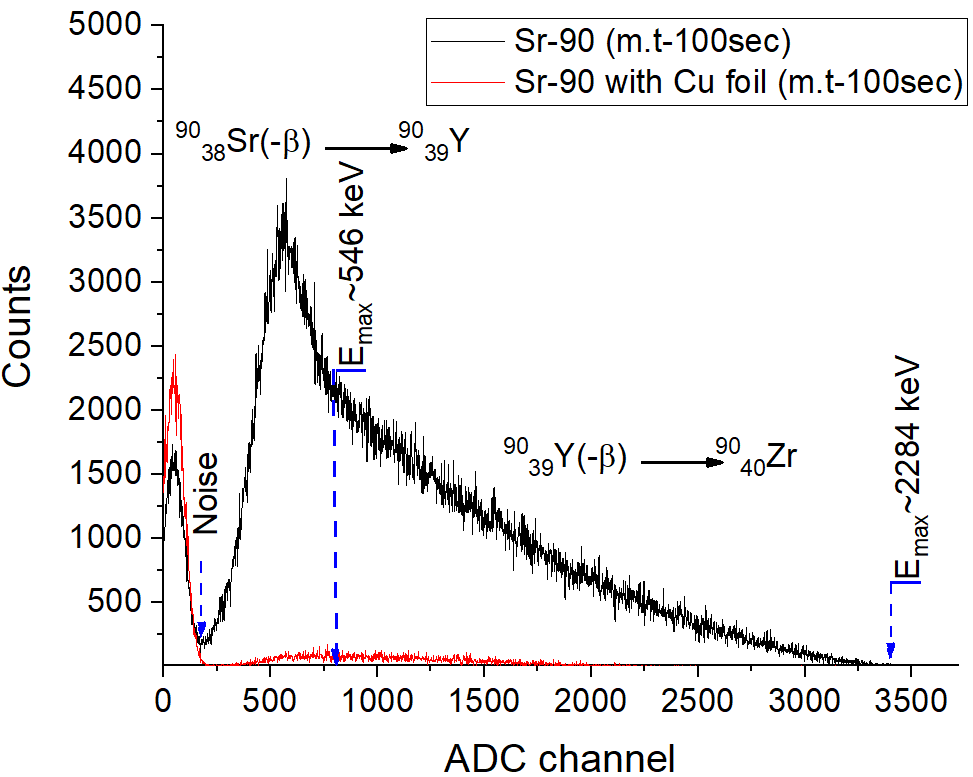}
\caption{Energy spectra of $^{90}$Sr measured with the SiPM array + plastic scintillators (the variable gain was selected 15 dB).}
\label{figure:6}
\end{figure}

Energy resolution of the monoenergetic electrons with an energy of 626 keV was measured to be 25$\%$ at 15 dB variable gain. Notably, the Compton edge of gamma rays was observed to increase with the energy of the incident gamma rays.
Figure \ref{figure:6} shows  the energy spectrum obtained from $^{90}$Sr. $^{90}$Sr undergoes decay, emitting an electron with a maximum energy of 546 keV, transforming into $^{90}$Y, which further decays, emitting an electron with a maximum energy of 2274 keV, and finally  resulting in $^{90}$Zr \cite {bib26}. When a copper foil was placed between the scintillator and the source, the beta particles were completely absorbed by the foil, and only gamma and x-ray events were observed in the spectrum. The obtained results were found  to be in agreement with the data reported in other works \cite {bib26}. 

\begin{figure}[tp]
\centering
\includegraphics[width=.5\textwidth]{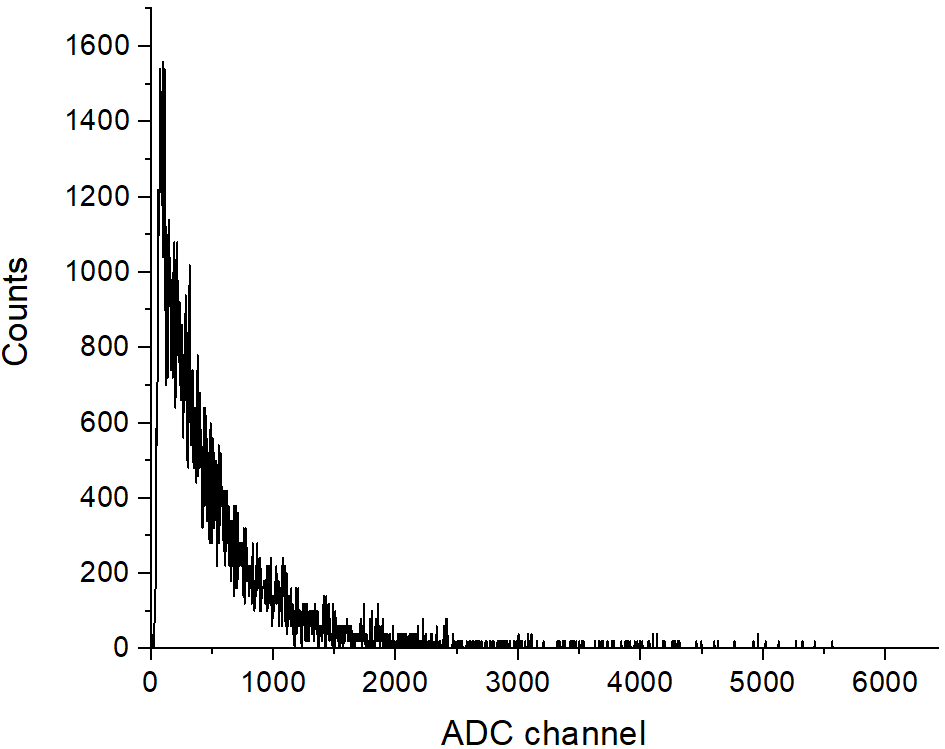}
\caption{Energy spectra of CmBe measured with the SiPM array + plastic scintillators (the variable gain was selected 15 dB).}
\label{figure:7}
\end{figure}

CmBe neutron sources were utilized to test the neutron detection performance of the plastic scintillator. The energy spectrum of CmBe is presented in figure \ref{figure:7}. The CmBe source emits both neutrons and gamma rays, both of which have energies within the range of up to 10000 keV.
The detection of fast neutrons is achieved through elastic collisions between neutrons and hydrogen atoms, leading to the generation of protons. The newly generated protons, resulting from the nuclear interaction with hydrogen, can be effectively detected using the scintillator. In order to minimize the effects of gamma rays, Pb shielding material was employed, which effectively attenuates gamma rays while allowing neutrons to penetrate and interact with the scintillator.

The source was placed inside Pb shielding box with a thickness of 5 cm, which effectively reduced  the intensity of gamma rays. This arrangement enabled the observation of the effect of fast neutrons on the plastic scintillator. However, due to the limited size of the scintillator, it was insufficient to fully detect the high-energy part of neutrons, resulting in only a few events being visible in the high-energy region of the spectra. The implementation of the pulse shape and zero crossing discrimination methods \cite {bib27} proved ineffective in distinguishing between neutrons and gamma rays. Despite efforts to utilize this technique, the response of the scintillator did not exhibit the required differentiation between the two types of radiation. As a result, the pulse shape and zero crossing discrimination methods were not effective for distinguishing neutrons from gamma rays in this particular scintillator setup. Further investigations and alternative methods may be necessary to achieve the desired discrimination capability in future studies. 

\section{\label{sec:level4} Conclusion}

The gamma ray, beta particle and neutron detection performance of the newly fabricated plastic scintillator by TENMAK was investigated. The  plastic scintillator exhibits two  absorption bands at 225 and 340 nm, with an absorption edge observed at 410 nm. The wavelength of maximum emission for the plastic scintillator is  $\sim$ 427 nm, with the light yield of  6134 photons for 1 MeV incident radiation. 
The obtained results showed that the Compton edge of detected gamma rays was observed to increase linearly with increasing the energy of gamma rays. The plastic scintillator demonstrated effective detection of beta particles emitted by $^{137}$Cs and $^{90}$Sr sources, with an energy resolution of 25 $\%$ for monoenergetic (626 keV) electrons. Furthermore, this type of plastic scintillator exhibited sensitivity to gamma rays and fast neutrons. The scintillator did not demonstrate the capability to discriminate between neutrons and gamma rays using the pulse shape discrimination method. Considering these advantageous characteristics, the combination of this plastic scintillator with SiPM technology makes it well-suited for detecting beta particles, gamma rays and neutrons in radiation monitoring and applications related to public safety. Further investigations are essential to attain optimal light yields and enhance the properties of the scintillator.

\section*{ACKNOWLEDGMENTS}

This project has received funding from the European Union's Horizon 2021 Research and Innovation Programme under the Marie Sklodowska-Curie grant agreement 101086178.






\end{document}